\documentclass[conference]{IEEEtran}
\IEEEoverridecommandlockouts
\usepackage{amsmath,amssymb,amsfonts}
\usepackage{algorithmic}
\usepackage{graphicx}
\usepackage{textcomp}
\usepackage{xcolor}
\usepackage{comment}
\usepackage{threeparttable}
\def\BibTeX{{\rm B\kern-.05em{\sc i\kern-.025em b}\kern-.08em
    T\kern-.1667em\lower.7ex\hbox{E}\kern-.125emX}}
    
\usepackage{subfig}

\usepackage[export]{adjustbox}

\usepackage[
backend=biber,
style=ieee,
sorting=ynt
]{biblatex}
\usepackage{etoolbox}
\makeatletter
\patchcmd{\@maketitle}
  {\addvspace{0.5\baselineskip}\egroup}
  {\addvspace{-2\baselineskip}\egroup}
  {}
  {}
\makeatother

\begin{document}

\title{Enhancing Virtual Distillation with Circuit Cutting for Quantum Error Mitigation\\
}

\makeatletter
\newcommand{\linebreakand}{%
  \end{@IEEEauthorhalign}
  \hfill\mbox{}\par
  \mbox{}\hfill\begin{@IEEEauthorhalign}
}
\makeatother 

\author{\IEEEauthorblockN{Peiyi Li}
\IEEEauthorblockA{\textit{NC State University} \\
Raleigh, NC, USA\\
pli11@ncsu.edu}
\and
\IEEEauthorblockN{Ji Liu}
\IEEEauthorblockA{\textit{Argonne National Laboratory} \\
Lemont, IL, USA \\
ji.liu@anl.gov}
\and
\IEEEauthorblockN{Hrushikesh Pramod Patil}
\IEEEauthorblockA{\textit{NC State University} \\
Raleigh, NC, USA \\
hpatil2@ncsu.edu}
\linebreakand 
\IEEEauthorblockN{Paul Hovland}
\IEEEauthorblockA{\textit{Argonne National Laboratory} \\
Lemont, IL, USA \\
hovland@mcs.anl.gov}
\and
\IEEEauthorblockN{Huiyang Zhou}
\IEEEauthorblockA{\textit{NC State University} \\
Raleigh, NC, USA \\
hzhou@ncsu.edu}
}
\maketitle

\begin{abstract}
Virtual distillation is a technique that aims to mitigate errors in noisy quantum computers. It works by preparing multiple copies of a noisy quantum state, bridging them through a circuit, and conducting measurements. As the number of copies increases, this process allows for the estimation of the expectation value with respect to a state that approaches the ideal pure state rapidly. However, virtual distillation faces a challenge in realistic scenarios: preparing multiple copies of a quantum state and bridging them through a circuit in a noisy quantum computer will significantly increase the circuit size and introduce excessive noise, which will degrade the performance of virtual distillation. To overcome this challenge, we propose an error mitigation strategy that uses circuit-cutting technology to cut the entire circuit into fragments. With this approach, the fragments responsible for generating the noisy quantum state can be executed on a noisy quantum device, while the remaining fragments are efficiently simulated on a noiseless classical simulator. By running each fragment circuit separately on quantum and classical devices and recombining their results, we can reduce the noise accumulation and enhance the effectiveness of the virtual distillation technique. Our strategy has good scalability in terms of both runtime and computational resources. We demonstrate our strategy's effectiveness through noisy simulation and experiments on a real quantum device.
\end{abstract}


\section{Introduction}
Quantum computation holds immense promise for solving complex problems beyond the reach of classical computers. However, quantum systems are highly susceptible to errors caused by various noise sources. 
These errors can severely degrade the performance and reliability of quantum algorithms. Therefore, it is essential to develop techniques that can mitigate the effects of errors and enhance the quality of quantum computation.

In the quest to mitigate errors and enhance the reliability of quantum computations, various techniques have been proposed. One notable technique, known as virtual distillation~\cite{Huggins2021,vikstål2022study} or error suppression by derangement~\cite{Koczor2021}, aims to achieve exponential suppression of errors when estimating the expectation value of an observable. 
The main idea behind virtual distillation is to prepare multiple copies of a noisy quantum state, denoted as $\rho$, \textquotedblleft bridging\textquotedblright~them through a circuit and then perform measurements on these copies~\cite{Koczor2021}. By utilizing the information obtained from these measurements, one can estimate the expectation value with respect to the state $\rho^M / \operatorname{Tr}\left(\rho^M\right)$, where $M$ represents the number of copies.
As $M$ increases, this state approaches the closest pure state to $\rho$ exponentially fast~\cite{Huggins2021}. By effectively approaching a pure state, the technique enables a more accurate estimation of the expectation value, thereby enhancing the reliability of quantum computations.

Virtual distillation holds great potential as a technique to mitigate the detrimental effects of noise in quantum systems. However, for near-term quantum devices, several obstacles can hinder the effectiveness of virtual distillation.
First, the need for 2-qubit gates, which are essential for \textquotedblleft bridging\textquotedblright~multiple copies of a quantum state, can lead to a significant increase in the number of required SWAP gates. Due to the limited qubit connectivity in near-term quantum devices, additional swap gates must be inserted before applying 2-qubit gates that involve non-adjacent qubits. Unfortunately, both these inserted SWAP gates and the 2-qubit gates used to \textquotedblleft bridge\textquotedblright~multiple state copies introduce significant noise, substantially compromising the reliability of the virtual distillation process.
Second, the process of preparing multiple copies of a circuit is vulnerable to crosstalk errors~\cite{Sarovar_2020}. Crosstalk between instructions can corrupt the quantum state when multiple instructions are executed simultaneously~\cite{Murali_2020}. Virtual distillation can be more susceptible to crosstalk between instructions, given that the preparation of multiple state copies significantly increases the likelihood of parallel gate executions on nearby qubits.
Furthermore, the virtual distillation circuit can be more susceptible to detection crosstalk (or readout) crosstalk~\cite{Sarovar_2020} because the preparation of multiple state copies and the subsequent measurements introduce a large number of measurement operations. As program size expands and the number of measurement operations increases, it becomes more susceptible to readout crosstalk~\cite{Das_2021}.

To improve the reliability and effectiveness of virtual distillation on near-term quantum devices, further noise mitigation techniques are needed. Reference~\cite{Koczor2021} has explored zero-noise extrapolation~\cite{Temme2017extrapolation,Li2017extrapolation} to mitigate the effects of noise of the virtual distillation circuits. However, they primarily focused on demonstrating the performance of zero-noise extrapolation in a depolarizing noise model. In a realistic scenario where various noise sources exist in the virtual distillation circuit, applying zero-noise extrapolation becomes challenging. The presence of multiple noise sources complicates the process of finding an accurate curve-fitting model, rendering zero-noise extrapolation less effective in practice.

This limitation highlights the need to develop alternative noise mitigation strategies that are better suited for real-world scenarios. 
In this paper, we propose noise mitigation using a quantum circuit-cutting strategy for virtual distillation on quantum devices. 
Our proposed approach involves cutting the virtual distillation circuit into smaller fragments and running them independently. By doing so, our proposed scheme can mitigate a significant amount of noise, and our experiments demonstrate the effectiveness of our strategy in enhancing the performance of virtual distillation on real quantum devices.

This paper is organized as follows: In Section~\ref{sec_vd}, we provide a review of the virtual distillation protocol and the circuit structure typically used in this protocol, then we analyze the overhead of a virtual distillation circuit on real devices and highlight the motivation for our work. Section~\ref{sec_cut} introduces our quantum circuit-cutting strategy and discusses its potential for enhancing virtual distillation. Section~\ref{sec_exp_setup} describes the experimental setup, including the benchmarks, evaluation metrics, and noise models. Experimental results and performance analysis will be presented in Section~\ref{sec_exp}. Finally, Section~\ref{sec_conclusion} concludes the paper.

\section{background and motivation}\label{sec_vd}
\subsection{Theory of Virtual Distillation}
This section reviews the theory of Virtual Distillation, and the notation used is based on Ref.~\cite{Huggins2021, vikstål2022study}.
When the state preparation process is affected by \emph{incoherent} noise, the desired pure quantum state, represented as $\left|\psi\right\rangle\left\langle\psi\right|$, can be distorted into a mixed state, described by the density matrix:
\begin{equation}\label{eq_mix_state}
\rho=\sum_{k=1}^d \lambda_k\left|\psi_k\right\rangle\left\langle\psi_k\right|
\end{equation}
In Equation (\ref{eq_mix_state}), $\lambda_k$ represents the probability of the system in the state $\left|\psi_k\right\rangle$, where $\lambda_k \leq 1$ and $\sum_k \lambda_k=1$. For simplicity, we assume that the first state $\left|\psi_1\right\rangle\left\langle\psi_1\right|$ in the mixture is the dominant eigenstate, with $\lambda_1$ being the dominant eigenvalue.

To mitigate the effects of noise and restore the purity of the state, the mixed state $\rho$ can be exponentiated $M$ times, normalizing $\rho^M$ yields:
\begin{equation}
\frac{\rho^M}{\operatorname{Tr}\left(\rho^M\right)}=\frac{\sum_{k=1}^d \lambda_k^M\left|\psi_k\right\rangle\left\langle\psi_k\right|}{\sum_{k=1}^d \lambda_k^M}
\end{equation}

The mitigated expectation value, denoted as $\langle O\rangle_{\text{mitigated}}$, is obtained by evaluating the trace of the observable $O$ with the state $\rho^M / \operatorname{Tr}\left(\rho^M\right)$: 
\begin{equation}
\langle O\rangle_{\text {mitigated }}:=\frac{\operatorname{Tr}\left(O \rho^M\right)}{\operatorname{Tr}\left(\rho^M\right)}
\end{equation}
By computing $\langle O\rangle_{\text{mitigated}}$, virtual distillation approximates the expectation value with respect to the dominant eigenstate $\left|\psi_1\right\rangle\left\langle\psi_1\right|$. Although the dominant eigenstate $\left|\psi_1\right\rangle\left\langle\psi_1\right|$ is not necessarily the desired state $\left|\psi\right\rangle\left\langle\psi\right|$, theoretical analysis~\cite{koczor2021dominant} shows the mismatch between these two states is exponentially smaller than the build-up of the other erroneous states $\left|\psi_k\right\rangle\left\langle\psi_k\right|_{k\neq1}$. This approach enables the mitigation of errors and provides an enhanced estimation of quantum observables.

\subsection{Circuit Implementation of Virtual Distillation}

To estimate the expectation value of the state $\rho^M / \operatorname{Tr}\left(\rho^M\right)$, Ref.~\cite{Huggins2021} utilize the following equation:
\begin{equation}
\frac{\operatorname{Tr}\left(O \rho^M\right)}{\operatorname{Tr}\left(\rho^M\right)}=\frac{\operatorname{Tr}\left(O^{\mathbf{i}} S^{(M)} \rho^{\otimes M}\right)}{\operatorname{Tr}\left(S^{(M)} \rho^{\otimes M}\right)}
\end{equation}
Here, $O^{\mathbf{i}}$ represents the observable $O$ acting on an arbitrary subsystem $i$, while $S^{(M)}$ denotes the cyclic shift operator applied to $M$ systems.

This equation provides an expression for estimating the desired expectation value by relating it to the trace of the observable $O^{\mathbf{i}} S^{(M)} \rho^{\otimes M}$ and the trace of $S^{(M)} \rho^{\otimes M}$. To estimate these two traces, it is necessary to prepare $M$ copies of the state $\rho$. For practical implementation on near-term quantum devices, we will focus on the case of preparing two copies of the state.

The circuit implementation involves two main steps. First, $M$ copies of the state $\rho$ need to be prepared. In the context of preparing two copies, it is represented as $\rho^{\otimes 2}$. The second step is to measure the expectation value of $\rho^{\otimes 2}$ with respect to the observables $S^{(M)}$ and $O^{\mathbf{i}} S^{(M)}$. Reference~\cite{Huggins2021} discusses two methods for measuring these observables. One approach involves the introduction of ancilla qubits, where the ancilla qubits are prepared in the state $|+\rangle=(|0\rangle+|1\rangle) / \sqrt{2}$ and a sequence of controlled-SWAP gates are applied to \textquotedblleft bridge\textquotedblright~the $M$ copies of $\rho$. 
Alternatively, another approach eliminates the requirement for ancilla qubits. This method relies on diagonalizing gates designed to diagonalize the observables $S^{(M)}$ and $O^{\mathbf{i}} S^{(M)}$. These diagonalizing gates \textquotedblleft bridge\textquotedblright~the $M$ copies of $\rho$ and allow measurement of these observables in the computational basis. We can then estimate the expectation values based on the measurement outcomes. The virtual distillation circuit with diagonalizing gate implementation is shown in Fig.~\ref{fig_vd_mapping}.
\begin{figure}[htbp]
\centerline{\includegraphics[width=\linewidth]{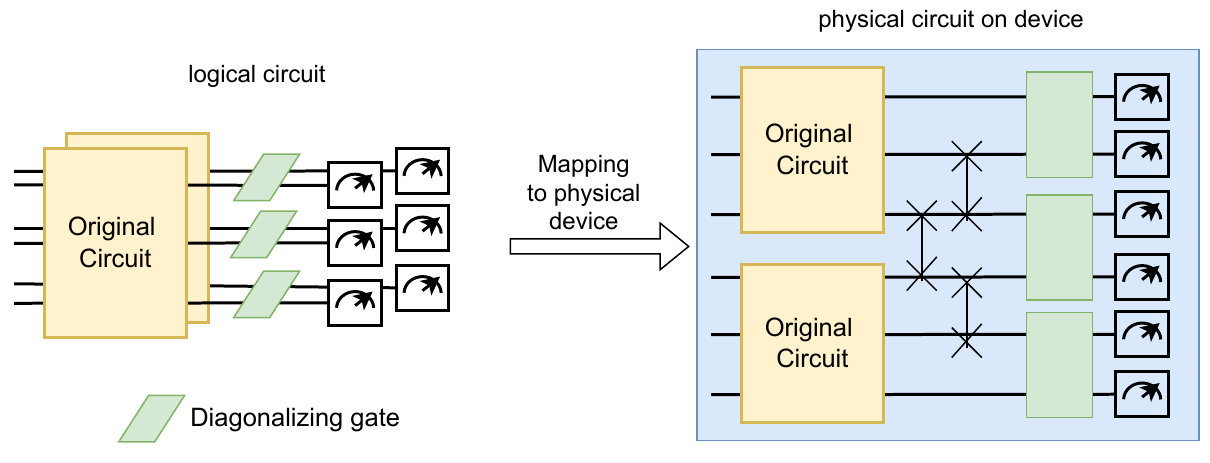}}
\caption{The virtual distillation circuit with two copies of the original circuit and the diagonalizing gates. When mapping the virtual distillation circuit to a device with limited connectivity, extra SWAP gates are introduced to move the diagonalizing gates to adjacent qubits with coupling.}
\label{fig_vd_mapping}
\end{figure}

In practical scenarios, the execution of controlled-SWAP gates on near-term quantum devices often introduces substantial noise, rendering the implementation of virtual distillation using ancilla qubits impractical. As a result, the remainder of this paper focuses on the virtual distillation implementation by utilizing diagonalizing gates. 

\subsection{Virtual Distillation Circuit Complexity on Device with Limited Connectivity}
\label{subsec:VD_on_device}
According to the findings in Ref.~\cite{Huggins2021}, the implementation of the virtual distillation circuit requires only a single additional layer of diagonalizing gates. Consequently, the number of extra gates needed for virtual distillation exhibits a linear growth relative to the number of qubits in the circuit. Furthermore, Ref.~\cite{Koczor2021} suggests that the preparation of the quantum state $\left|\psi\right\rangle$ typically necessitates $\mathcal{O}[a(N) N]$ gates, where $a(N)$ represents the computation depth. However, in scenarios where the computational problems extend beyond a constant-depth circuit, the gate count in the primary computation increases at a faster rate than $\mathcal{O}(N)$. Consequently, as the computation is scaled up, the additional gate count required for constructing the virtual distillation circuit becomes relatively less significant.

Nevertheless, it is crucial to acknowledge that the aforementioned study does not take into account the inherent connectivity limitations of near-term quantum devices. The qubits in the superconducting quantum computers~\cite{Google72qubit, ibm127qubit} and the neutral atom array quantum computers~\cite{graham2022neutral_atom} are not fully connected. While the existing trapped-ion devices featuring tens of qubits offer full connectivity, the challenge lies in designing a scalable QCCD (Quantum Charge-Coupled Device) architecture~\cite{pino2021QCCD} that comprises thousands of qubits. Such an architecture would consist of multiple fully-connected clusters; however, the inter-connection across these clusters remains limited. The connectivity limitation in the quantum devices requires the insertion of additional SWAP gates before applying the diagonalizing gates. Fig.~\ref{fig_vd_mapping} shows an example of mapping a 6-qubit virtual distillation circuit to a physical device with linear connectivity. Three SWAP gates are inserted for qubit routing. In the worst-case scenario, each diagonalizing gate requires at most $N-1$ SWAP gates to move the two logical qubits to adjacent physical qubits. Therefore, the upper bound of the number of extra gates introduced by virtual distillation is $\mathcal{O}(N^2)$. As the circuit size grows, the number of these additional gates increases at a rate faster than linear, resulting in a substantial overhead.

In order to assess the impact of these additional SWAP gates in the virtual distillation circuit, we employed a RealAmplitudes circuit from Qiskit library~\cite{realamplitudes}. 
The RealAmplitudes circuit is commonly utilized as an ansatz circuit in chemistry applications and consists of alternating layers of rotation-Y gates and CNOT gates. An example of the RealAmplitudes circuit can be seen in Fig.~\ref{fig_real_amplitudes_circuit}.
\begin{figure}[htbp]
\centerline{\includegraphics[width=\linewidth]{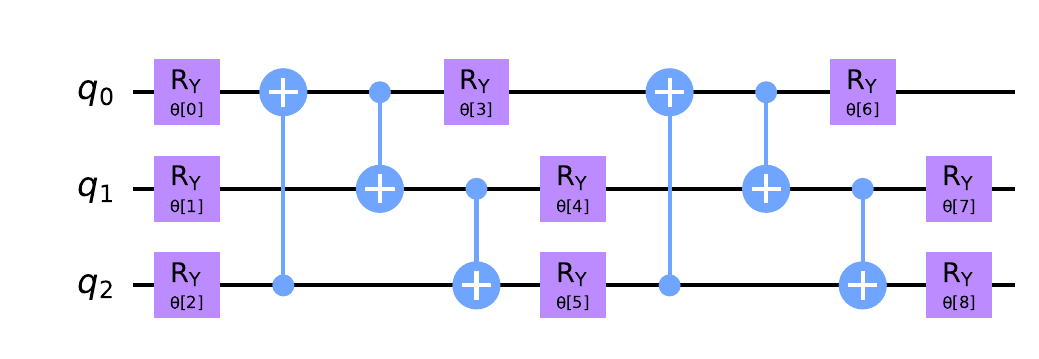}}
\caption{Example of a 3 qubit RealAmplitudes circuit with 2 repetitions and circular entanglement.}
\label{fig_real_amplitudes_circuit}
\end{figure}

\begin{figure}
\centering
\subfloat[Circuit transpiled using coupling map of fully connectivity\label{fig_cnot_counts_fully_connectivity}]{
\begin{adjustbox}{width=0.45\linewidth}
\includegraphics[width=\linewidth]{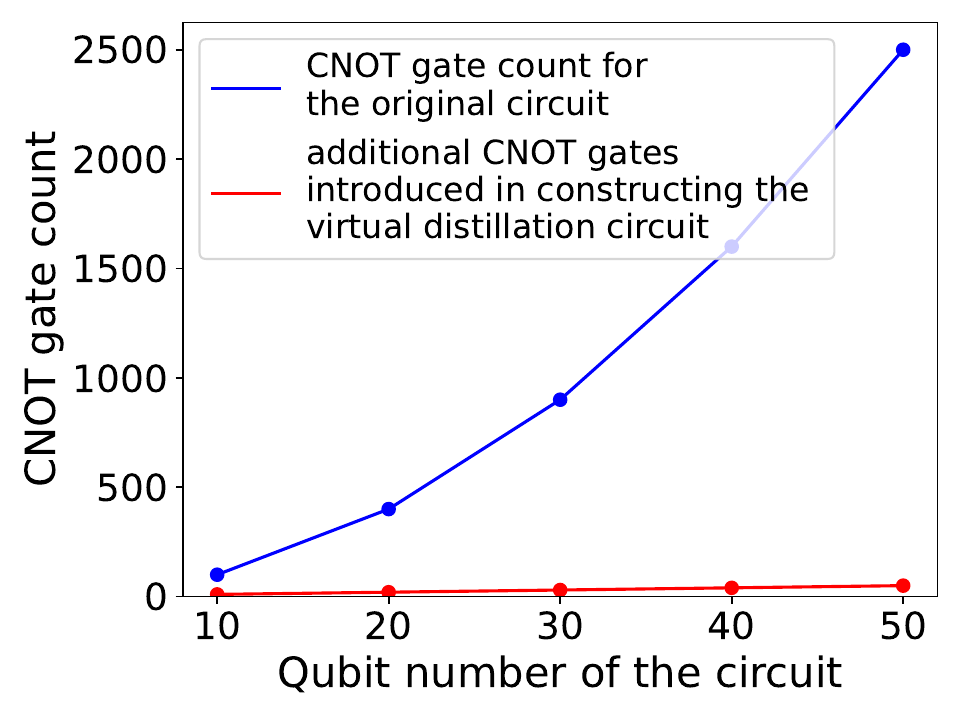}
\end{adjustbox}
}\hfil
\subfloat[Circuit transpiled using coupling map from \texttt{ibm\_sherbrooke}\label{fig_cnot_counts}]{
\begin{adjustbox}{width=0.45\linewidth}
\includegraphics[width=\linewidth]{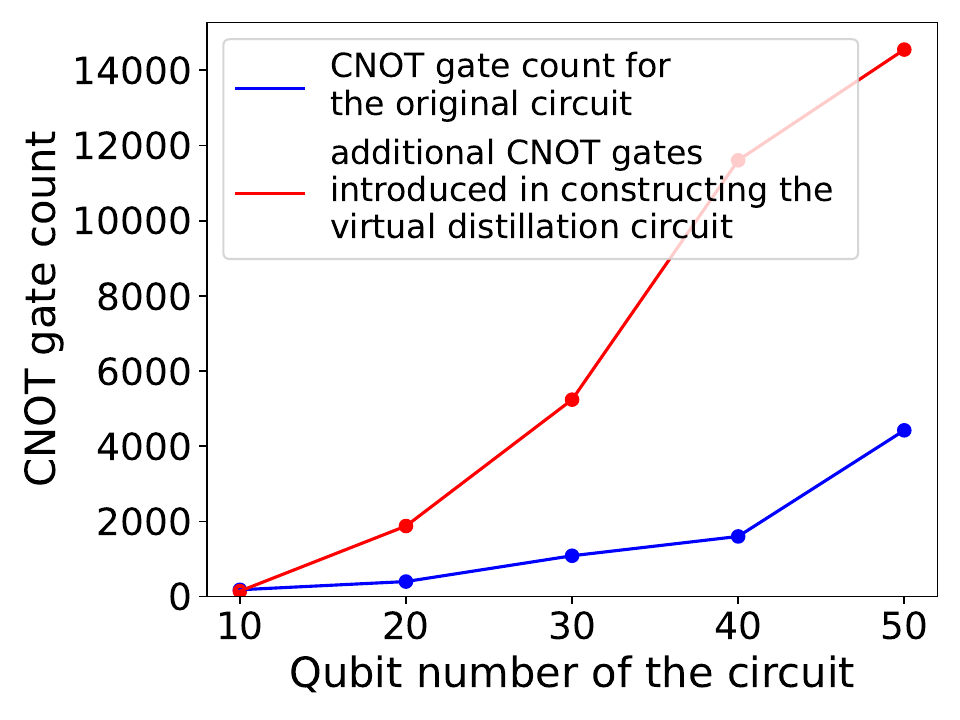}
\end{adjustbox}
}\hfil
\caption{Change of circuit size with respect to the number of qubits. Note that the total CNOT gate count in the virtual distillation circuit is equal to two times the CNOT gate count in the original circuit, plus the additional CNOT gates introduced during the construction of the diagonalizing gates.}
\label{fig_cnot_counts_total}
\end{figure}

In our study, we utilized the RealAmplitudes circuit as the original circuit for state preparation, varying the number of qubits and layers. Subsequently, we applied the virtual distillation approach to examine the impact of additional SWAP gates on circuit size.
To evaluate the overhead associated with applying virtual distillation on a near-term device, we transpiled both the original circuits and the circuits after applying the virtual distillation approach. The transpilation process utilized a coupling map derived from the 127-qubit IBM machine, named \texttt{ibm\_sherbrooke}. This machine adopts a heavy hex lattice topology commonly used in IBM quantum machines. The optimization level is set to 3 to minimize the number of additional inserted SWAP operations.
The resulting CNOT gate counts after transpilation are presented in Fig. \ref{fig_cnot_counts}. 
For comparison purposes, we also included the circuit translation results using a fully connected coupling map in Fig. \ref{fig_cnot_counts_fully_connectivity}.

We systematically increased both the number of qubits (ranging from 10 to 50) and the alternating layers of rotation-Y gates and CNOT gates (ranging from 10 to 50) to modify the gate complexity of the original circuit.
Figure \ref{fig_cnot_counts_fully_connectivity} illustrates the ideal scenario with a fully connected coupling map, where the additional gate count required for constructing a virtual distillation circuit grows linearly with respect to the number of qubits, resulting in a relatively smaller overhead compared to the increase in gate count of the original circuit. However, by visualizing the changes in circuit size on a limited connected coupling map from the real quantum device through Fig. \ref{fig_cnot_counts}, it becomes evident that as the circuit scales up, the number of additional gates required for constructing the virtual distillation circuit remains substantial, introducing significant overhead. This observation highlights the fact that running the virtual distillation circuit on a limited connectivity device significantly increases circuit complexity, thereby undermining the effectiveness of the virtual distillation approach. In light of this, we propose a solution for enabling the application of the virtual distillation approach on near-term quantum devices by employing circuit-cutting techniques to reduce gate complexity significantly. 

\section{enhancing virtual distillation }\label{sec_cut}
In this section, we will delve into enhancing the virtual distillation circuit by utilizing quantum circuit-cutting techniques. We begin by introducing the theory behind quantum circuit cutting. Subsequently, we explain how this method can be applied to the virtual distillation circuit, highlighting the advantages of our approach.

\subsection{Quantum Circuit Cutting}
This section provides a basic overview of Quantum Circuit Cutting~\cite{Peng_2020,perlin2021quantum,GideonUchehara2022}.  
In Ref.~\cite{perlin2021quantum}, it was shown that an arbitrary quantum state represented by a density matrix $\rho$, can be decomposed as:
\begin{equation}\label{eq_state_decompose}
\rho \simeq \frac{1}{2} \sum_{M \in \mathcal{B}} M \otimes \operatorname{tr}_n\left(M_n \rho\right)
\end{equation}
Here, $\mathcal{B}$ represents the basis of self-adjoint $2 \times 2$ matrices, where for convenience, we can select $\mathcal{B} \equiv \{X, Y, Z, I\}$ as the basis. The symbol $\operatorname{tr}_n$ denotes a partial trace operation with respect to qubit $n$, while $M_n$ denotes the action of the operator $M$ on qubit $n$, with the operator $I$ acting on the remaining qubits. 

\begin{figure}[htbp]
\centerline{\includegraphics[width=\linewidth]{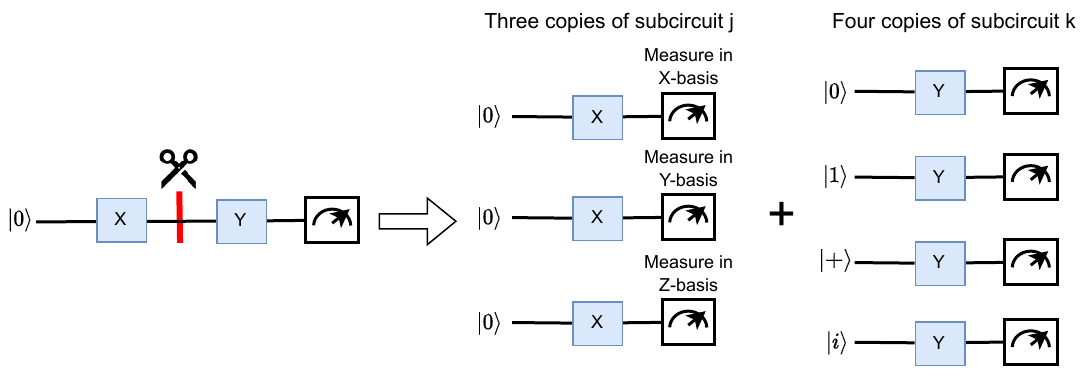}}
\caption{Circuit cutting for a single-qubit circuit.}
\label{fig_cutting_demo} 
\end{figure}

Each term in the equation (\ref{eq_state_decompose}) can be divided into two components.
The first component, $\operatorname{tr}_n\left(M_n \rho\right)$, corresponds to the measurement of the observable $M_n$ while the system is in the state $\rho$. This portion of the circuit can be referred to as subcircuit $j$. The second component involves the initialization or preparation of the eigenstates of $M$. This segment can be denoted as subcircuit $k$.
By following this approach, the equation demonstrates how a quantum state can be reconstructed after a cut is made on one of its qubits, as illustrated in Fig.~\ref{fig_cutting_demo}. This technique forms the core of quantum circuit cutting.

As shown in Fig.~\ref{fig_cutting_demo}, cutting one wire results in three copies of the subcircuit $j$ and four copies of the subcircuit $k$. It is important to highlight that the total number of copies exhibits exponential growth as the number of cuts made to a circuit increases. Therefore, an efficient scheme with circuit-cutting needs to find good cutting points that limit the total number of cuts in the circuit. While identifying suitable cutting points in general poses challenges, the inherent structure of the virtual distillation circuit facilitates the identification of good cutting points.

\subsection{Applying Quantum Circuit Cutting in Virtual Distillation}

\begin{figure*}[htbp]
\centerline{\includegraphics[width=\linewidth]{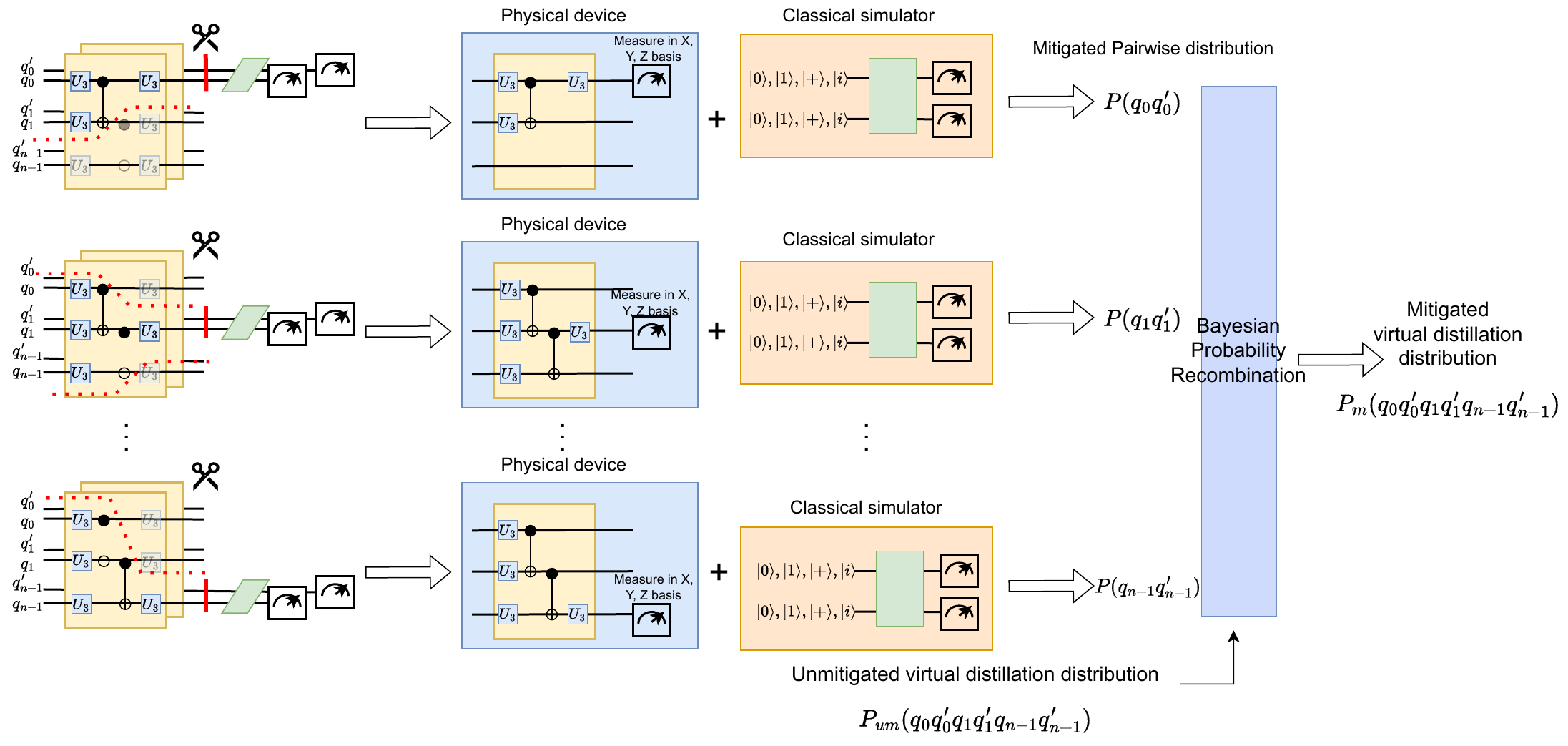}}
\caption{Demonstration of applying quantum circuit-cutting to the virtual distillation circuit. The transparent gates in the circuit are the ones that have no dependency on the diagonalizing gates.}
\label{fig_vd_cutting}
\end{figure*}

As discussed in Section~\ref{subsec:VD_on_device}, implementing virtual distillation circuits on real devices has two challenges: 1) Additional SWAP gates are inserted before applying the diagonalizing gates. 2) Both the inserted SWAP gates and the diagonalizing gates are subject to noise and may hinder the virtual distillation result. 

These two challenges can be effectively addressed by cutting the virtual distillation circuit into fragments of the diagonalizing gates and fragments of the original circuit that prepares the noisy quantum state. First, by independently executing these smaller fragments, the requirement for inserting SWAP operations is significantly reduced. Second, since the size of each diagonalizing gate does not increase with the number of qubits in the original circuit, we can efficiently simulate the diagonalizing gates on a noise-free classical simulator. This simulation aids in mitigating the noise introduced by the diagonalizing gates and enhances the overall performance. Liu et al. introduced the Simulated Quantum Error Mitigation (SQEM) framework~\cite{liu2022classical} where they leveraged circuit cutting to simulate the Pauli Check Sandwiching circuit~\cite{gonzales2023PCS} for quantum error mitigation. 

Ideally, our objective is to cut out all the diagonalizing gates simultaneously; however, the number of subcircuit copies grows exponentially with the number of cuts. To overcome this limitation, we leverage the fact that diagonalizing gates are only applied pairwise and propose a pairwise circuit cutting scheme. The proposed scheme is demonstrated in Fig.~\ref{fig_vd_cutting}. The original circuit for generating the noisy quantum state is an n-qubit circuit and we prepare two copies of the noisy state in the virtual distillation circuit. 
\begin{itemize}
\item First, the virtual distillation circuit is executed on a physical device to acquire the unmitigated noisy output distribution $P_{um}(q_0q_0'q_1q_1'...q_{n-1}q_{n-1}')$. 
\item Then, we will replicate the virtual distillation circuit n times to obtain n mitigated pairwise distributions $P(q_0q_0),...,P(q_{n-1}q_{n-1}')$. Since the goal is to obtain pairwise distribution, each replica only consists of gates that have dependency with the measurement, i.e., two identical subcircuits of the original circuit and a diagonalizing gate. The diagonalizing gate is cut out and simulated on a noise-free simulator, while the rest of the circuit is executed on a physical device.
\item Lastly, to obtain the final mitigated result, we need to update the unmitigated noisy output distribution $P_{um}(q_0q_0'q_1q_1'...q_{n-1}q_{n-1}')$ with mitigated pairwise distributions $P(q_0q_0),...,P(q_{n-1}q_{n-1}')$. We utilize the recombination method described in~\cite{liu2022classical} to merge these results and produce the overall mitigated output.
\end{itemize}

To optimize the circuit and reduce noise, we utilize the following optimizations in the scheme. Firstly, when executing the part on the classical simulator, we can reuse the results of the diagonalizing gate if the separated diagonalizing gates have the same matrix. This approach saves computation time and resources. Secondly, for the part executed on the quantum device, we eliminate any gates that are not predecessors of the separated diagonalizing gate. This pruning process results in a more compact and less noisy circuit that is customized for the specific quantum device. As shown in Fig.~\ref{fig_vd_cutting}, the transparent gates in the diagram are the ones that have no dependency on the diagonalizing gates and can be eliminated.

\subsection{Advantage of Applying Quantum Circuit Cutting in Virtual Distillation}

The application of the quantum circuit cutting technique in virtual distillation offers several notable advantages. These advantages contribute to enhancing the performance and effectiveness of the virtual distillation circuit.

First, by dividing the virtual distillation circuit into multiple parallel steps, we enable efficient execution of the circuit. This parallelization allows us to take advantage of the available computational resources and accelerate the overall processing time. 
Furthermore, by reusing the results of the diagonalizing gate on the classical simulator, we reduce the computational overhead associated with redundant calculations.  Additionally, by removing irrelevant gates that are not predecessors of the separated diagonalizing gate, we eliminate unnecessary operations, resulting in a more concise and streamlined circuit. This pruning process mitigates the impact of noise and significantly improves the quality of the mitigated pairwise distributions $P(q_0q_0),...,P(q_{n-1}q_{n-1}')$. 
Also, by eliminating non-essential gates, we create a circuit that is optimized for the connectivity and operational constraints of the quantum device. This customization improves the circuit's compatibility with the device, increasing the likelihood of successful execution and achieving desired results.

To obtain the final mitigated output, the outcomes from each step, which correspond to individual pairs of qubits in the circuit, need to be combined. The recombination method described in~\cite{liu2022classical} provides an effective approach for merging these results. By leveraging classical post-processing techniques, we can integrate the outcomes from each step and generate the overall mitigated output.

In summary, applying the quantum circuit cutting technique in virtual distillation brings several advantages, including efficient parallel execution, noise reduction, and accurate outcome recombination. These advantages contribute to improving the performance, reliability, and scalability of virtual distillation, making it a promising approach for near-term quantum devices.

\subsection{Complexity Analysis}\label{sec_complexity}
The proposed scheme is scalable with respect to the number of qubits $N$ in the original circuit. The total time is directly proportional to the number of runs performed on the physical device.  As we replicate the virtual distillation circuit $N$ times and the circuit cutting results in three copies for each replica, the total number of hardware runs is $\mathcal{O}(3N)$. The runtime can be reduced since the hardware execution of these copies can be parallelized.  

Regarding the computational complexity associated with simulating diagonalizing gates, the size of the diagonalizing gate is determined by the number of copies $M$ in the virtual distillation setup. Since a large value of $M$ significantly increases the size of the virtual distillation circuit, it is common practice to set $M$ to 2, and the diagonalizing gates become two-qubit gates. Consequently, the classical simulation complexity is bounded by a constant value $\mathcal{O}(C)$. Since the number of diagonalizing gates in the circuit is $N$ and the gates are simulated independently, the classical computational complexity is $\mathcal{O}(N)$. Notably, many diagonalizing gates are identical so we can reuse the simulation results to reduce the classical computational overhead.

\section{Experimental Methodology}~\label{sec_exp_setup}
\vspace*{-8mm}
\subsection{Benchmark}
To evaluate the effectiveness of our approach, we conducted experiments using the Variational Quantum Eigensolver (VQE) algorithm to solve the MaxCut problem. We utilized the RealAmplitudes circuit as Ansatz for the VQE algorithm, which consists of alternating rotation Y gates and CNOT gates, e.g.,  Fig.~\ref{fig_real_amplitudes_circuit}.
In our experiments, we set the alternating layer number of the rotation Y gates and CNOT gates to 2. The circuit parameters were fixed to the optimal values obtained using the `COBYLA' optimizer on a noise-free simulator. The problem Hamiltonian for the MaxCut problem is defined as:
\begin{equation}
H=\sum_{(i, j) \in E} \frac{1}{2}\left(1-Z_i Z_j\right)
\end{equation}
where $Z_i$ is the Pauli Z operator acting on qubit $i$. $E$ is the set of the graph edges in the MaxCut problem. 

\subsection{Evaluation Criteria}
To assess the effectiveness of different error mitigation approaches, we calculate the absolute error of the expectation value for the problem Hamiltonian compared to the noise-free expectation value. This allows us to quantify the deviation of each error mitigation approach from the noise-free solution.
To assess the gate complexity of different error mitigation approaches, we transpile the circuit into basis gates and count the number of CNOT gates. This provides a measure of the gate operations required by each error mitigation approach, offering a comparison of the circuit complexities.

\subsection{Experiment Platform}
Our experiments were conducted using the Qiskit framework v0.39.0. 
We utilized both a real quantum machine, a $27$-qubit \texttt{ibm\_hanoi}, and a simulator, Qiskit Aer simulator, in our experiment. The simulator allows us to simulate quantum circuits under both noise-free and noisy conditions. 
We set the shot number to 10,000 for both the simulator and the real quantum machine. 

\subsection{Noise Models}\label{sec_noise_model}

Qiskit offers noise models that incorporate certain sources of noise from real quantum devices, and we use the noise model from real quantum device \texttt{ibm\_hanoi} as our basic noise model. This basic noise model considers factors such as gate errors, gate time, $T_1$ and $T_2$ relaxation times, and readout errors for each qubit. The parameter of this noise model is based on the calibration data taken on September 27, 2023. The median CNOT error is  $7.936e-3$, the median gate time is $346.667\,\text{n}\text{s}$, the median readout error is $1.200e-2$, the median T1 is $120.385\,\mu\text{s}$, and the median T2 is $138.652\,\mu\text{s}$.

In order to provide a more realistic evaluation of our approach, we incorporated additional noise sources in our experiments. Specifically, we modeled the ZZ crosstalk for CNOT gates and the readout crosstalk.  
By including these noise sources, we aimed to accurately capture the impact they have on the performance and effectiveness of different error mitigation approaches.

To model the ZZ crosstalk for CNOT gates, we adopted a method similar to the one described in~\cite{Ahsan2022crosstalk}. This involved introducing additional $R_{ZZ}$ gates in the circuit whenever there were CNOT gates in the same layer and in adjacent positions. We set the angle $\theta$ of the inserted $R_{ZZ}$ gate to $-\pi / 3.5$ based on the Ref.~\cite{Ahsan2022crosstalk}.

Readout crosstalk refers to the phenomenon where the measurement of one qubit can be affected by the state of neighboring qubits due to unintended coupling~\cite{Sarovar_2020}.
The basic noise model only considers the single-qubit readout errors which assumes that the readout noise acts independently on each individual qubit. To model the readout crosstalk, we introduced a 2-qubit readout error matrix for pairs of neighboring qubits. 
In our simulations, we set the readout error matrix as: 
\[
\begin{bmatrix}0.991 & 0.003 & 0.003 & 0.003\\ 0.003 & 0.991 & 0.003 & 0.003\\ 0.003 & 0.003 & 0.991 & 0.003\\ 0.003 & 0.003 & 0.003 & 0.991\end{bmatrix}
\]

\subsection{Comparison with Extrapolation Approach}
We conducted a comparison with the zero-noise extrapolation approach~\cite{Temme2017extrapolation, Li2017extrapolation} commonly employed for quantum error mitigation. The extrapolation technique aims to estimate the expectation value of an observable by extrapolating measurements obtained from circuits at different noise levels. 

In the case of the extrapolation approach for virtual distillation, we followed a similar methodology as described in Ref.~\cite{Koczor2021}, where we scaled up the diagonalizing gates in the virtual distillation circuit for the extrapolation process. This scaling was specifically applied to the diagonalizing gates, as they are often subject to noise and can limit the precision of the virtual distillation approach in realistic scenarios. 

In our experiments, we choose three scale factors 1, 3, and 5, for linear extrapolation.

The implementation of our approach and the comparison results with the extrapolation approach are publicly available at https://github.com/peiyi1/project\_error\_mitigation.
\section{evaluation}~\label{sec_exp} 
In this section, we compare different error mitigation approaches under different noise models. Then we run experiments on real machines to show that our approach works for real quantum devices.

\subsection{Comparison under Different Noise Models}

To evaluate the noise robustness of our circuit-cutting approach, we conducted a comparative analysis with different error mitigation approaches across various noise models. The results of these comparisons are presented in Table~\ref{table:result_4qubit} and Table~\ref{table:result_6qubit}, which display the simulation outcomes for the 4-qubit and 6-qubit VQE circuits, respectively.

To showcase the circuit optimization achieved through our approach, we compared the number of CNOT gates in different error mitigation techniques. Our focus on the CNOT gate count is motivated by the fact that 2-qubit gates typically introduce more gate noise in comparison to single-qubit gates and the CNOT gate is a widely used 2-qubit gate in superconducting systems. After applying the circuit-cutting technique, the CNOT gate count decreases compared with other mitigation approaches. By eliminating unnecessary gates, our approach reduces the overall gate count, thereby mitigating the impact of noise. This reduction in gate count leads to improved pairwise distributions, ultimately enhancing the accuracy of the output distributions in the presence of noise.

Table~\ref{table:result_4qubit} and Table~\ref{table:result_6qubit} also provide the expectation value and the absolute error obtained by different approaches under various noise models. 
The results demonstrate that as the complexity of the noise model increases, the error mitigation effectiveness of the virtual distillation approach diminishes. However, our circuit-cutting approach maintains a high level of error mitigation across different noise models. This achievement is attributed to the division of the circuit into smaller fragments, which greatly reduces the occurrence of CNOT gates executing at the same layer and in adjacent positions, thereby eliminating a significant portion of the ZZ crosstalk noise. Table~\ref{table:result_4qubit} and Table~\ref{table:result_6qubit} display the count of $R_{ZZ}$ gates when modeling the ZZ crosstalk for CNOT gates, indicating that our circuit-cutting approach introduces less ZZ crosstalk, whereas the original virtual distillation approach introduces a substantial amount of ZZ crosstalk. 
Furthermore, the tables also illustrate the efficient mitigation of readout crosstalk achieved by our circuit-cutting approach. This is achieved by reducing the number of measurement operations in the circuit.  
Conversely, the original virtual distillation circuit requires the preparation of at least two copies of the circuit, resulting in an amplification of readout crosstalk due to the increased number of qubits and measurement operations.

When comparing our circuit-cutting approach with the extrapolation approach, we observe that for smaller virtual distillation circuits, as illustrated in Table~\ref{table:result_4qubit}, the extrapolation method indeed exhibits effective error mitigation. However, as shown in Table~\ref{table:result_6qubit}, as the circuit size increases and the complexity of the noise model grows, the effectiveness of the extrapolation approach decreases. This is because the presence of multiple noise sources complicates the establishment of an accurate curve-fitting model for extrapolation. In contrast, our circuit-cutting approach consistently achieves error reduction even in the presence of more complex noise sources. This highlights the effectiveness and reliability of our approach in mitigating errors in quantum circuits.

\begin{table*}[bhtp]
  \caption{Simulation results for the 4-qubit VQE circuit}
  \label{table:result_4qubit}
  \centering
    \resizebox{\linewidth}{!}{%
    \begin{threeparttable}
  \begin{tabular}{|c|c|c|c|c|c|c|c|c|}
  \hline
   \multicolumn{1}{|c|}{ } & \multicolumn{2}{c|}{ gate count} &  \multicolumn{3}{c|}{expectation value} &  \multicolumn{3}{c|}{absolute error}\\
    \hline
     error mitigation method & $CNOT$ gate count  & $R_{ZZ}$ gate count    & basic\tnote{1} & basic+GCT \tnote{2}& basic+GCT+RCT \tnote{3} & basic & basic+GCT &basic+GCT+RCT\\\hline
     no mitigation  & 17         & 0        & -2.659\tnote{4} & -2.656 & -2.637    & 0.313 & 0.316 & 0.335 \\\hline
virtual distillation   & 63         & 16       & -2.792\tnote{5} & -2.696 & -2.681    & 0.180 & 0.276 & 0.291 \\\hline
virtual distillation + extrapolation  & 63,71,79\tnote{6} & 16,24,32    & -3.003 & -2.911 & -2.895     & 0.031 & 0.061 & 0.077 \\\hline
virtual distillation + circuit cutting & 11,14,17,17\tnote{7}& 0,0,0,0  & -2.914 & -2.914 & -2.913     & 0.058 & 0.058 & 0.059 \\\hline

  \end{tabular}
  \begin{tablenotes}
       \item [1] basic denotes the basic noise model that Qiskit offers that incorporates noise from real quantum device ibm\_hanoi
       \item [2] basic + GCT denotes the basic noise model, which includes the ZZ crosstalk for CNOT gates
       \item [3] basic + GCT + RCT denotes the basic noise model that includes both the ZZ crosstalk for CNOT gates and the readout crosstalk 
       \item [4] The ideal expectation value for the original circuit in a noise-free simulator is -2.972
       \item [5] The expectation value for the virtual distillation circuit with noise-free diagonalizing gates is -2.965
       \item [6] The numbers 63, 71, and 79 represent the CNOT gate count for the extrapolation circuit with noise scales of 1, 3, and 5, respectively.
       \item [7] The numbers 11,14,17 and 17 represent the CNOT gate counts for the virtual distillation circuit when cutting and measuring each pair of qubits in the circuit, respectively.
       
     \end{tablenotes}
  \end{threeparttable}
  }
\end{table*}

\begin{table*}[bhtp]
  \caption{Simulation results for the 6-qubit VQE circuit}
  \label{table:result_6qubit}
  \centering
    \resizebox{\linewidth}{!}{%
    \begin{threeparttable}
  \begin{tabular}{|c|c|c|c|c|c|c|c|c|}
  \hline
   \multicolumn{1}{|c|}{ } & \multicolumn{2}{c|}{ gate count} &  \multicolumn{3}{c|}{expectation value} &  \multicolumn{3}{c|}{absolute error}\\
    \hline
     error mitigation method & $CNOT$ gate count   & $R_{ZZ}$ gate count    & basic & basic+GCT & basic+GCT+RCT & basic & basic+GCT & basic+GCT+RCT\\\hline
     no mitigation  & 31         &  0       & -4.278 \tnote{1} & -4.300 & -4.269                      & 0.644 & 0.622 & 0.653 \\\hline
virtual distillation   & 117         &  36     & -4.562\tnote{2} & -4.222 & -4.199                    & 0.360 & 0.699 & 0.723 \\\hline
virtual distillation+ extrapolation  & 117,129,141 &   36,56,72  & -4.914 & -4.506 & -4.482           & 0.008 & 0.416 & 0.440 \\\hline
virtual distillation + circuit cutting & 16,18,21,26,28,31&    2,0,0,8,4,0  & -4.842 & -4.785 & -4.783& 0.080 & 0.137 & 0.139 \\\hline

  \end{tabular}
  \begin{tablenotes}
       \item [1] The ideal expectation value for the original circuit in a noise-free simulator is -4.922
       \item [2] The expectation value for the virtual distillation circuit with noise-free diagonalizing gates is -4.903
     \end{tablenotes}
  \end{threeparttable}
  }
\end{table*}

\subsection{Real Device Results}
To validate the practical applicability of our approach, we conducted experiments on a 27-qubit quantum device \texttt{ibm\_hanoi}. The results for a 4-qubit and a 6-qubit VQE circuit are presented in Table~\ref{table_real_device}. Our circuit-cutting approach outperforms other error mitigation methods, highlighting the effectiveness of our approach in real-world quantum computing environments.

In addition to the experiments reported in Table~\ref{table_real_device}, we also performed experiments on a 10-qubit VQE circuit. However, despite applying circuit cutting, the resulting subcircuits still contained a substantial number of gates, leading to significant noise that prevented us from obtaining adequately mitigated pairwise distributions required for updating the noisy output distribution effectively. Consequently, we were unable to achieve satisfactory results. This further emphasizes the scalability and practical limitations of existing error mitigation approaches for larger quantum circuits.

\begin{table}[]
\centering
\caption{Real device results for the VQE circuit}
\label{table_real_device}
\resizebox{\linewidth}{!}{%
\begin{threeparttable}
\begin{tabular}{|c|c|c|c|c|}
\hline
\multicolumn{1}{|c|}{ } & \multicolumn{2}{c|}{ 4-qubit circuit} &  \multicolumn{2}{c|}{ 6-qubit circuit}  \\ \hline
\begin{tabular}[c]{@{}c@{}}error mitigation\\ method\end{tabular} & expectation value & absolute error & expectation value & absolute error\\ \hline
no mitigation                             & -2.540 \tnote{1} & 0.432  & -3.635 \tnote{2} &1.287 \\ \hline
virtual distillation                           & -1.964  & 1.008  & -2.659  &2.263 \\ \hline
\begin{tabular}[c]{@{}c@{}}virtual distillation\\ + extrapolation\end{tabular}    & -2.101 & 0.871 & -3.002  &1.920 \\ \hline
\begin{tabular}[c]{@{}c@{}}virtual distillation\\ + circuit cutting\end{tabular}  & -2.747 & 0.225  & -4.498  &0.424 \\ \hline
\end{tabular}
 \begin{tablenotes}
       \item [1] The ideal expectation value for the original circuit in a noise-free simulator is -2.972
       \item [2] The ideal expectation value for the original circuit in a noise-free simulator is -4.922
     \end{tablenotes}
  \end{threeparttable}
}
\end{table}

\section{conclusion}~\label{sec_conclusion}
In this paper, we propose a circuit-cutting based scheme aimed at enhancing the virtual distillation technique on near-term quantum devices. We observe the virtual distillation's performance is hampered by the limited connectivity and the noisy operations in the near-term quantum devices. To address these challenges, we propose an efficient and effective scheme that involves cutting the virtual distillation circuit into fragments and simulating the diagonalizing gates on noise-free simulators. This approach allows us to obtain high-quality pairwise distributions that can be utilized to update the original noisy distribution. Our experiments on a noisy simulator and real device show that our proposed approach outperforms both the canonical and the extrapolation-enhanced virtual distillation methods.

\section*{Acknowledgements}
We thank the anonymous reviewers for their valuable comments. This work is partly funded by NSF grants 1818914 (with a subcontract to NC State University from Duke University), and 2120757 (with a subcontract to NC State University from the University of Maryland). It is also supported by the U.S. Department of Energy, Office of Science, National Quantum Information Science Research Centers.
\printbibliography

\end{document}